# MODIFYING THE ENTITY RELATIONSHIP MODELING NOTATION: TOWARDS HIGH QUALITY RELATIONAL DATABASES FROM BETTER NOTATED ER MODELS

Dhammika Pieris, dhammika.pieris@monash.edu

School of Information Technology, Monash University

## *ABSTRACT*

*The entity relationship modelling using the original ER notation has been applauded providing a natural view of data in conceptual modelling of information systems. However, the current ER to relational model transformation algorithm is known to be insufficient in providing a complete and accurate representation of the ER model undertaken for transformation. In an effort to derive better transformations from ER models, we have understood that modifications should be introduced to both of the existing transformation algorithm as well as to the ER notation. Introducing some new concepts, we have adapted the original ER notation and developed a new transformation algorithm based on the existing one. This paper presents the modified ER notation with an ER diagram drawn based on the new notation.*

## 1. INTRODUCTION

This working paper proposes several modifications to the original ER modelling notation. The effort will help create ER diagrams (ERD) which are improved notation-wise and are of less confused, and which will in turn help design high quality relational database(Codd, 1970) models. Based on the modified ER modelling notation, the present ER to relational model transformation algorithm (Batini, Ceri, & Navathe, 1992; Ramez Elmasri & Navathe, 2007; R Elmasri & Navathe, 2011) will also be modified in order to obtain high quality database models from ERDs.

This paper will also be used as a supplementary document for empirical evaluations of the that investigates ways of designing high quality databases from ERDs. By a high quality database we presume a database semantically clear, complete, and accurate with respect to its predecessor ERD and is easy to reverse back to the ERD without any intuition.

To achieve the goal, an entire approach will be provided that consists of following components:

1. A modified notation scheme for ER modelling
2. A new ER to relational forward transformation algorithm
3. A criterion for assessing the quality of relational database schema (RDS)
4. A RDS to ER re-engineering algorithm
5. Applications

Accordingly, this paper presents the first component: "A modified notation scheme for ER modelling". The rest of the components will be published soon.

Transformation algorithms are inadequate(Pieris & Rajapakse, 2012), but it is not the only reason for a poor database designs to be resulted. Many methodologies and notation schemes are available for ER modelling(Batini et al., 1992; Ramez Elmasri & Navathe, 2007; R Elmasri & Navathe, 2011; Song, Evans, & Park, 1995; Teorey, Yang, & Fry, 1986). However, we take the view that most of them hardly can model a successfully transferable ER diagram to the relational model. To our understanding, the reason is the existence of a poor relationship among the concepts in the ER notation, the ER modelling method, the transformation algorithm, and the relational model. The first step to solve the issue is to provide a solid ER notation and modelling methodology that can create a transferable ERD as we expected.

## 2. A MODIFIED NOTATION SCHEEME FOR ER MODELING

The basic building blocks of the following notation system are based on the similar material presented in the sources:(Chen, 1976) and (Ramez Elmasri & Navathe, 2007; R Elmasri & Navathe, 2011). Using pairs of stars-"**"-we have encompassed the modifications we proposing. Anything non-encompassed should be considered borrowed from the above mentioned sources as a means of completing the proposing approach. Note that we have touched only few areas of the existing approach that we are most concerned with.

We claim that if an ER diagram is drawn obeying the proposing steps, it would be much easier to obtain a straight forward, semantically complete, interpretable and reversible relational database design by a new transformation algorithm to be presented soon. Following is the modified notation.

### *2.1. REPRESENTING REGULAR ENTITY TYPES*

1. The meaning of the regular or strong entity type will be further extended. Two conditions must be satisfied, for an entity type to be a regular entity type as follows.

I. The entity type must have at least one key attribute of its own(A key attribute is an attribute that help uniquely identify an instance of the entity type). **The key attribute should be in its atomic non divisible level**. For instance, a multi-valued attribute should not be used as a key attribute as the values of it might be further divisible.

II. **A key attribute must be a single attribute, and it should not be a combination of two or more attributes where each attribute in the combination partially contribute to form the key attribute**. For example, assume an entity type EmployeeBankAccount that has attributes: EmpNo, AccNo, EmployeeName, BankName, and AccountBalance, etc.. If so definitely, EmpNo and AccNo must both be combined together to uniquely identify an instance of this entity type. Thus, such an entity type is not considered to be a regular entity type. Note :- It is assumed that the AccNo is just a code of digits, and two Banks may have the same code for AccNo.

2. There may be multiple key attributes for an entity type
3. A regular entity type is represented by a rectangular box enclosing its name, for example, Employee, Department, and Manager in the bellow ERD, Figure 1.
4. **Names of different entity types must be different**. There cannot be two entity types with the same name.

### 2.2. REPRESENTING SUBTYPES

1. A subtype is defined to be a sub set of entity instances of an entity type. For example assume that there are a significant number of "Managers" among employees in a certain firm and the firm requires it's Managers to be separately portrayed in the ERD that describes the firm's data. Thus, there must be a subtype named "Manager" of the entity type "Employee" in the firm's ERD. Further a subtype must have following qualities.

    I.    **It must have at least one intrinsic or mutual property. **.

    II.    It must be represented by a rectangular box enclosing its name.

2. A subtype is connected to its supertype by a "Is-A" relationship arrow directed from the subtype to the supertype, for example, Manager is a subtype of the supertype Employee, and hence it is connected to the supertype by an arrow directed from the subtype Manager to the supertype.

## 2.3. REPRESENTING WEAK ENTITY TYPES

1. A weak entity type does not have its own key attributes, but it may have partial key attribute(s). A partial key attribute of a weak entity type combines with the key attribute of the owner entity type of the weak entity type to uniquely identify the instances of the weak entity type. A partial key attribute is denoted by underlining its name by a discrete line, for example, "Name" of the entity type "Dependent" which has been underlined using a discrete line. A weak entity type is denoted by double lined rectangle, for example, Dependent.

2. A weak entity type must have an owner entity type to which the weak entity type is to be connected by a weak entity relationship type, a double lined diamond shape that encloses the name of the weak-entity-relationship-type. For example, in the ERD, the double lined rectangle "DepandentOf" should be a weak entity relationship type that connects "Dependent" to its owner "Employee".

## 2.4. REPRESENTING ATTRIBUTES

1. A key attribute attached to an entity type is represented by underlining its name, for example, DepNo and Name which are the key attributes of the entity type Department are underlined.

2. **The name of at least one key attribute or the most desired key attribute of a regular entity type must be uniquely related to the name of its corresponding entity type."

    I. **To achieve this requirement, we propose re-naming the most desired key attribute (if more than one key attribute are existed) as follows. Select the most desired key attribute of a regular entity type and concatenate the first 3 letters of the name of the entity type to the name of the attribute in it's left side. For example, in the ERD, the key attributes: EmpNo of Employee, DepNo of the Department, and the ProNo of the Project have already been named as proposed**.

    II. **If two regular entity types have their names different but equal in first 3 letters, then the number of selecting letters from the names of the respective entity types should be increased from 3 to 4 or any other desired minimal number as appropriate. For example, assume that "Employee" and "Empowerment" are two regular entity types in an ERD, and assume just for demonstration purpose that each of them have the key attributes: "Ssn" and "SeqNo"; "Ssn" denote the "social security number", and "SeqNo" denote something called "sequence number". However the first three letters, "Emp", of both of the entity types are the same. Thus, it is suggested selecting the first four letters, "Empl" of the entity type "Employee" and the

"Empo" of the entity type "Empowerment". Thus, the names of the key attributes would then become EmplSsn and EmpoSeqNo respectively.

  III. **In a situation where a regular entity type contains multiple key attributes, it is recommended using the most desired key attribute for the above mentioned re-naming**.

3. We propose releasing the long argued requirement of naming each attribute of an entity type related to the name of the entity type, for example, Employee.Name, Employee.Address etc.
4. An attribute is represented by an oval enclosing its name and is attached to its entity type by a straight line, as per the attributes: "EmpNo" and "Name" are attached to the entity type "Employee".
5. A multi-valued attribute is represented by a double lined oval, for example, the attribute Location attached to Department.

## 2.5. COMMON RULES OF NAMING ENTITIES, ATTRIBUTES AND RELATIONSHIP TYPES

1. **Words used in the names of attributes, entity types and relationship types should be of capital letter initialized followed by simple letters, for example, Employee, Project and Department. Nouns must be in the singular form, so that we do not use the term "Locations" as the name of a multi-valued attribute. Instead we propose using the term "Location" **.
2. **All the attribute names and entity type names should be kept free from underscores, hyphens, and slashes or any other symbol but alphabet-letters only**.

3. **If multiple words are occurred in a name, they must be concatenated removing intermediate spaces, for example, the attribute name "StartDate" attached to the relationship type "Manages", is a concatenation of two words: "Start" and "Date"**

### 2.6. REPRESENTING RELATIONSHIP TYPES

1. A relationship type is denoted by a diamond that encloses the name of the relationship type, for example, the relationship type "Manages" that connect the entities: Manager and Department.
2. A relationship type that associates two entity types is called a binary relationship, for example, the relationship type "Assigned" defined between the entity types Employee and Department. Similarly, a relationship type that associates three entity types is called a ternary relationship type, and that associates n , n >2 entity types is called an n-array relationship type.

### 2.7. REPRESENTING CONSTRAINTS ASSOCIATED WITH RELATIONSHIP TYPES

1. In the ERD, "Assigned" is a relationship type existed between the regular entity types Employee and Department. The pairs of values (1, 1) and (4, 12) are called the *min-max-cardinality-ratio-constraints* of the participation of each of the entity types Employee and Department in the relationship type "Assigned" respectively. The values in a pair can also be denoted in general as (*min, max*), so that the values in a pair near to a particular entity type indicate that an instance of the entity type participate in at least *min* and at most *max* number of relationship type instances. For example the values in the pair (4, 12) indicates that an instance of the entity type Department (in other wards a particular department) participate at least 4 and at most 12 instances of the relationship type. The real world meaning of it is a department can deploy minimum 4 and maximum 12 employees.

2. The pair of *min* and *max* values of a cardinality ratio constraint is expected to lie in the range: *min ≤ max*, *max ≥ 1*

   A similar cardinality ratio: (1, 1) is defined in between "Employee" and "Assigned". It can be interpreted that a particular employee is assigned to at least 1 and at most 1 department. This relationship type is called a one-to-many (1:N) relationship type from Department to Employee, so that a particular department can be assigned with many employees. The Employee entity is said to be at the N-side of the relationship.

3. Thus, in a one-to-many relationship, the entity type near to the cardinality ratio that has the max value 1 is said to be at the N-side of the relationship.

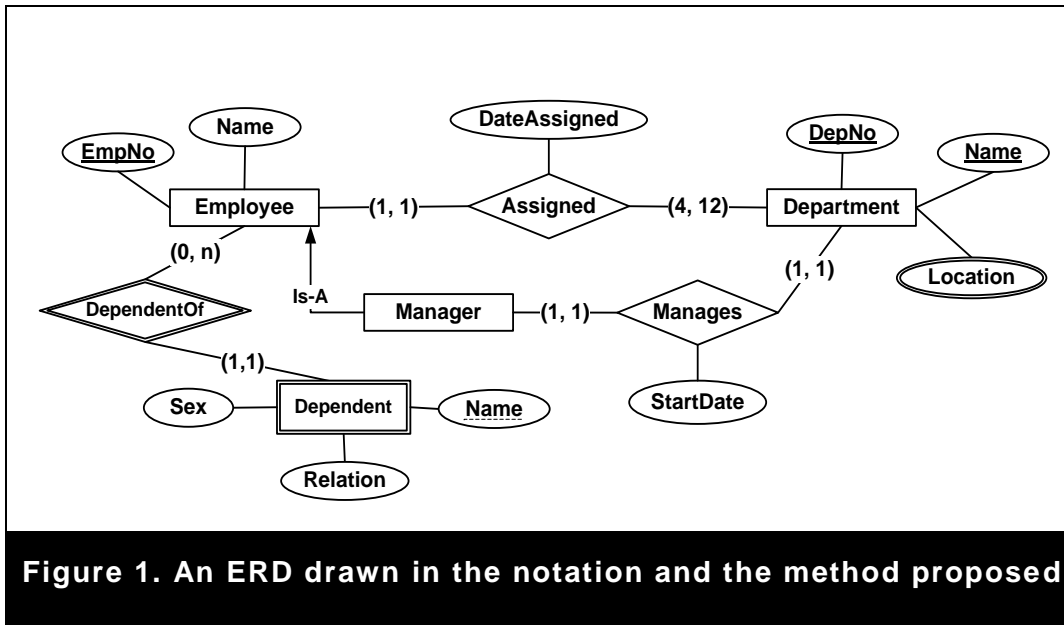

**Figure 1. An ERD drawn in the notation and the method proposed**

## 3. CONCLUSION

Any ERD willing to be transformed using the new algorithm to be proposed must first be modelled conforming to the above requirements, or if the ERD is already drawn, it must be re-adjusted accordingly. The ERD must be a conformed to the above rules in order to obtain a high quality RDS according to the quality criterion which is also to be presented soon.

We further believe a quality wise better database schema can be obtained even by applying the existing transformation algorithm on an ERD conformed to the above rules.

*Those who interested are invited to evaluate the material presented and send feedback to the author.*

Author's contacts: dhammika.pieris@monash.edu, dhammikapieris@yahoo.com